\documentclass[useAMS,a4,usenatbib]{mn2e}
\input psfig.sty

\def \xoff {\ifmmode x_{\rm off} \else $x_{\rm off}$ \fi}
\def \rhorms {\ifmmode \rho_{\rm rms} \else $\rho_{\rm rms}$ \fi}

\def \apj  {ApJ}
\def \apjl  {ApJL}
\def \apjs  {ApJS}
\def \prd {Phy.Rev.D}
\def \mnras {MNRAS}
\def \etal {et~al.~}
\def \chisq  {\ifmmode  \chi^2   \else  $\chi^2$  \fi}  
\def \chisqr {\ifmmode \chi^2_{\rm r} \else $\chi^2_{\rm r}$ \fi}
\def \spose#1{\hbox  to 0pt{#1\hss}}  
\def \lta{\mathrel{\spose{\lower 3pt\hbox{$\sim$}}\raise  2.0pt\hbox{$<$}}}
\def \gta{\mathrel{\spose{\lower  3pt\hbox{$\sim$}}\raise 2.0pt\hbox{$>$}}}
 
\def \ha  {\ifmmode H\alpha \else H$\alpha $ \fi}

\def \morgana {{\sc morgana}}

\def \kms {\ifmmode  \,\rm km\,s^{-1} \else $\,\rm km\,s^{-1}  $ \fi }
\def \kpc {\ifmmode  {\rm kpc}  \else ${\rm  kpc}$ \fi  }  
\def \Msun {\ifmmode M_{\odot} \else $M_{\odot}$ \fi} 
\def \hMsun {\ifmmode h^{-1}\,\rm M_{\odot} \else $h^{-1}\,\rm M_{\odot}$ \fi}
\def \hhMsun {\ifmmode h^{-2}\,\rm M_{\odot}\else $h^{-2}\,\rm M_{\odot}$ \fi}
\def \Lsun {\ifmmode L_{\odot} \else $L_{\odot}$ \fi} 
\def \hhLsun {\ifmmode h^{-2}\,\rm L_{\odot} \else $h^{-2}\,\rm L_{\odot}$ \fi}

\def \LCDM {\ifmmode \Lambda{\rm CDM} \else $\Lambda{\rm CDM}$ \fi}
\def \sig8 {\ifmmode \sigma_8 \else $\sigma_8$ \fi} 
\def \OmegaM {\ifmmode \Omega_{\rm M} \else $\Omega_{\rm M}$ \fi} 
\def \OmegaL {\ifmmode \Omega_{\rm \Lambda} \else $\Omega_{\rm \Lambda}$\fi} 
\def \Deltavir {\ifmmode \Delta_{\rm vir} \else $\Delta_{\rm vir}$ \fi}

\def \rs {\ifmmode r_{\rm s} \else $r_{\rm s}$ \fi} 
\def \rrm2 {\ifmmode r_{-2} \else $r_{-2}$ \fi} 
\def \ccm2 {\ifmmode c_{-2} \else$c_{-2}$ \fi} 
\def \cvir {\ifmmode c_{\rm vir} \else $c_{\rm vir}$ \fi} 
\def \cbar {\ifmmode \overline{c} \else $\overline{c}$ \fi}

\def \R200 {\ifmmode R_{200} \else $R_{200}$ \fi} 
\def \Rvir {\ifmmode R_{\rm vir} \else $R_{\rm vir}$ \fi}

\def \v200 {\ifmmode V_{200} \else $V_{200}$ \fi} 
\def \Vvir {\ifmmode V_{\rm  vir} \else  $V_{\rm vir}$  \fi} 
\def  \Vhalo  {\ifmmode V_{\rm halo} \else $V_{\rm halo}$ \fi}

\def \M200 {\ifmmode M_{200} \else $M_{200}$ \fi} 
\def \Mvir {\ifmmode M_{\rm  vir} \else $M_{\rm  vir}$ \fi}  
\def \Mshell  {\ifmmode M_{\rm shell} \else $M_{\rm shell}$ \fi}

\def \Nvir {\ifmmode N_{\rm  vir} \else $N_{\rm  vir}$ \fi}  

\def \Jvir {\ifmmode J_{\rm vir} \else $J_{\rm vir}$ \fi} 
\def \Jshell {\ifmmode J_{\rm shell} \else $J_{\rm shell}$ \fi}

\def \Evir {\ifmmode E_{\rm vir} \else $E_{\rm vir}$ \fi} 

\def \lam {\ifmmode \lambda  \else $\lambda$ \fi} 
\def \lamp {\ifmmode \lambda^{\prime} \else $\lambda^{\prime}$  \fi} 
\def \lampc {\ifmmode \lambda^{\prime}_{\rm c} \else
  $\lambda^{\prime}_{\rm c}$  \fi} 
\def \lambar {\ifmmode \bar{\lambda}  \else  $\bar{\lambda}$  \fi}  
\def  \lampbar  {\ifmmode \bar{\lambda^{\prime}} \else
  $\bar{\lambda^{\prime}}$\fi} 
\def \siglam {\ifmmode \sigma_{\lambda} \else $\sigma_{\lambda}$ \fi} 
\def \siglamp {\ifmmode                \sigma_{\lambda^{\prime}} \else
$\sigma_{\lambda^{\prime}}$\fi}

\def \Rd {\ifmmode R_{\rm d} \else $R_{\rm d}$ \fi} 
\def \Rs {\ifmmode R_{\rm s} \else $R_{\rm s}$ \fi}  
\def \Rd {\ifmmode R_{\rm d} \else $R_{\rm d}$ \fi}  
\def \Rcool  {\ifmmode R_{\rm  cool}  \else $R_{\rm cool}$ \fi} 
\def \RIII {\ifmmode  3.2\Rs \else $3.2\Rs$ \fi} 
\def \RII {\ifmmode 2.2\Rs \else $2.2\Rs$  \fi} 
\def \Reff {\ifmmode R_{\rm eff} \else $R_{\rm  eff}$ \fi} 
\def  \rb {\ifmmode r_{\rm b}  \else $r_{\rm b}$ \fi}

\def  \Sigmacrit   {\ifmmode  \Sigma_{\rm  crit}   
\else  $\Sigma_{\rm crit}$\fi} 
\def \Sig0 {\ifmmode \Sigma_{0} \else $\Sigma_{0}$ \fi}

\def \muI {\ifmmode \mu_{0,I} \else $\mu_{0,I}$ \fi}

\def \mgal {\ifmmode m_{\rm gal} \else $m_{\rm gal}$ \fi} 
\def \md {\ifmmode m_{\rm d} \else $m_{\rm d}$ \fi} 
\def \ms {\ifmmode m_{\rm   s}   \else   $m_{\rm   s}$   \fi}   
\def   \mdbar   {\ifmmode {\overline{m}}_{\rm d} \else
  ${\overline{m}}_{\rm d}$ \fi} 
\def \msbar {\ifmmode  \bar{m}_{\rm  s}  \else  $\bar{m}_{\rm s}$
  \fi}  
\def  \Md {\ifmmode M_{\rm d}  \else $M_{\rm d}$ \fi} 
\def  \Ms {\ifmmode M_{\rm s} \else $M_{\rm  s}$ \fi} 
\def \Mb {\ifmmode  M_{\rm b} \else $M_{\rm b}$ \fi} 
\def \Mstar {\ifmmode  M_{\rm star} \else $M_{\rm star}$ \fi}
\def \Mdisc {\ifmmode M_{\rm disc} \else $M_{\rm disc}$ \fi}

\def \Jd {\ifmmode J_{\rm d} \else $J_{\rm d}$ \fi} 
\def \Jb {\ifmmode J_{\rm b} \else $J_{\rm b}$ \fi}  
\def \fb {\ifmmode  f_{\rm b} \else $f_{\rm b}$ \fi}

\def  \jd  {\ifmmode j_{\rm  d}  \else  $j_{\rm  d}$ \fi}  
\def  \jdmd {\ifmmode \frac{j_{\rm  d}}{m_{\rm d}} \else
  $\frac{j_{\rm d}}{m_{\rm d}}$ \fi} 
\def \fj {\ifmmode f_{\rm j} \else $f_{\rm j}$ \fi} 
\def \ft {\ifmmode f_{\rm t}  \else $f_{\rm t}$ \fi} 
\def  \fM {\ifmmode f_{\rm M} \else $f_{\rm M}$ \fi}

\def  \Vd {\ifmmode  V_{\rm  d}  \else $V_{\rm  d}$  \fi} 
\def  \Vcool {\ifmmode V_{\rm cool} \else $V_{\rm cool}$ \fi} 
\def \Vcirc {\ifmmode V_{\rm circ}  \else $V_{\rm circ}$  \fi} 
\def \VIII  {\ifmmode V_{3.2} \else $V_{3.2}$ \fi} 
\def  \VII {\ifmmode V_{2.2} \else $V_{2.2}$ \fi}
\def \Vobs {\ifmmode V_{\rm obs}  \else $V_{\rm obs}$ \fi} 
\def \Vdisc {\ifmmode V_{\rm disc} \else  $V_{\rm disc}$ \fi} 
\def \Vmax {\ifmmode V_{\rm  max} \else  $V_{\rm max}$  \fi} 
\def  \Vmaxobs{\ifmmode V_{\rm max}^{\rm obs}\else  $V_{\rm max}^{\rm
    obs}$\fi}  
\def \Vtot {\ifmmode V_{\rm tot} \else $V_{\rm tot}$  \fi} 
\def \Vrot {\ifmmode V_{\rm rot} \else  $V_{\rm rot}$  \fi} 
\def  \Vflat {\ifmmode  V_{\rm  flat} \else $V_{\rm flat}$ \fi}

\def \Ups {\ifmmode \Upsilon  \else $\Upsilon$ \fi} 
\def \YB {\ifmmode \Upsilon_B \else $\Upsilon_B$ \fi} 
\def \YI {\ifmmode  \Upsilon_I  \else $\Upsilon_I$ \fi} 
\def \DeltaIMF {\ifmmode \Delta_{\rm IMF} \else $\Delta_{\rm IMF}$ \fi}

\def\LCDM{$\Lambda$CDM }

\def\c200{$c_{200}$}

\title[DM nature and Milky Way Satellites] {How cold is Dark Matter?
Constraints from Milky Way Satellites}
\author[A.V. Macci\`o \& F. Fontanot]  {Andrea
  V. Macci\`o$^{1}$\thanks{maccio@mpia.de} \& Fabio Fontanot$^{2,1}$ \\ 
$^1$Max-Planck-Institut f\"ur Astronomie, K\"onigstuhl 17, 69117
 Heidelberg, Germany \\ 
$^2$INAF-Osservatorio Astronomico, Via Tiepolo 11, I-34131 Trieste,
  Italy \\ }

\begin{document}
             
\date{submitted to MNRAS}
             

\maketitle           

\label{firstpage}
             
\begin{abstract}
We test the luminosity function of Milky Way satellites as a constrain
for the nature of Dark Matter particles. We perform dissipationless
high-resolution $N$-body simulations of the evolution of Galaxy-sized
halo in the standard Cold Dark Matter (CDM) model and in four Warm
Dark Matter (WDM) scenarios, with a different choice for the WDM
particle mass ($m_w$). We then combine the results of the numerical
simulations with semi-analytic models for galaxy formation, to infer
the properties of the satellite population. Quite surprisingly we find
that even WDM models with relatively low $m_{w}$ values (2-5 keV) are able
to reproduce the observed abundance of ultra faint ($M_v<-9$) dwarf
galaxies, as well as the observed relation between Luminosity and mass
within 300 pc. Our results suggest a lower limit of $1$ keV for thermal warm
dark matter, in broad agreement with previous results from
other astrophysical observations like Lyman-$\alpha$ forest and
gravitational lensing.

\end{abstract}


\setcounter{footnote}{1}

\section{Introduction}
\label{sec:intro}

The inflationary cold dark matter scenario gives a clear prediction
for the initial fluctuation spectrum responsible for the formation of
the Large Scale Structure in the Universe, with considerable power
down to very small scales. As a consequence we expect the mass
function of dark matter haloes to rise steeply towards low
masses. N-body simulations have indeed revealed that in Cold Dark
Matter (CDM) models, all Galaxy-sized haloes ($M_{DM} \sim 10^{12}
M_\odot$) should contain a large number of embedded subhaloes that
survive the collapse and virialization of the parent structure (e.g
Springel \etal 2008). Although the predicted number of substructures
is in reasonable agreement with observed luminosity functions (LFs) in
cluster sized haloes, in Milky Way (MW) sized haloes the number of
predicted sub-haloes exceeds the number of observed satellites by at
least an order of magnitude (Klypin \etal 1999, Moore \etal 1999).
This tension between models and observations represents, together
with the detailed reproduction of the star formation histories 
(e.g. Salvadori \etal 2008) and mass profiles of local dwarf galaxies
(e.g. Walker \etal 2009), 
one of the most relevant questions for the present day theories 
of galaxy formation and evolution.

A possible cosmological solution to this discrepancy is to replace
cold dark matter with a warm species (WDM, Bode, Ostriker \& Turok
2001 and references therein). The warm component acts to reduce the
small-scale power, resulting in fewer galactic subhaloes and lower
halo central densities (Col{\'{\i}}n \etal 2000,2008). One possible WDM
candidate is a sterile neutrino which exhibit a significant primordial
velocity distribution and thus damp primordial inhomogeneities on
small scales (e.g. Hansen \etal 2002, Abazajian \& Koushiappas 2006
Boyarsky, Ruchayskiy, Shaposhnikov 2009).
Since ground based experiments may be far from directly studying in 
detail the nature and properties of the actual DM particle, it is of fundamental
importance to determine astrophysical constraints on the maximum free
streaming length of any warm candidate.

Limits on the mass of dark matter particles can be obtained from several 
astrophysical observations: theoretical phase space density studies combined 
with observations of stellar dynamics in Milly Way satellites (e.g.
Boyanovsky, de Vega, Sanchez 2008; de Vega Sanchez \& Sanchez 2009),
luminosity function of high redshift QSOs (Song \& Lee 2009), abundance of 
dwarf galaxies in the Local Volume (Zavala \etal 2009) and the size 
of (min)-voids around the Local Group (Tikhovov \etal 2009).
Perhaps one of the most powerful tool for constraining the
matter power spectrum are Lyman-$\alpha$ forest observations 
(neutral hydrogen absorption in the
spectra of distant quasars, Narayanan \etal 2000, Viel \etal 2005).
Lyman-$\alpha$ observations allows the possibility to studying the
power spectrum down to small scales and over a large range of redshifts. 
Using HIRES data, a lower limit of $m_{w} = 1.2$ keV has
been reported by Viel \etal (2008). Even tighter
limits can be obtained combining Lyman-$\alpha$ observations with SDSS 
results (see Boyarsky \etal 2009a and references therein). 
These results have been recently confirmed by Miranda \&
Macci\`o (2007) in an independent estimation of lower limits for the
mass of the WDM particle based on QSO lensing observations.

In the last five years our knowledge on the number and properties of
satellite galaxies around our the MW has tremendously increased thanks
to the results from the Sloan Digital Sky Survey (SDSS). The
homogeneous sky coverage of the SDSS provided the first determination
of the volume corrected MW satellite luminosity function down to
luminosities as faint as 100 $L_{\odot}$ (Koposov \etal
2008). Exploiting this advancement in the observational knowledge of
our own Galaxy, several theoretical works have revised the problem of
satellite number density, coming to the conclusion that it is possible
to reconcile the observational evidences with the predictions of the
standard (L)CDM scenario (e.g. Koposov \etal 2009, Macci\`o \etal
2009, Li \etal 2009a).

The aim of the present work is to expand the results presented in
Macci\`o \etal 2010 (M10, hereafter), trying to use a combination of
observational data and theoretical predictions to infer significant
constraints on the allowed mass of any warm dark matter particle.
We start from the findings of M10 and we re-simulate one of the DM haloes 
presented in that work, in a WDM scenario for different choices of $m_w$. 
We then combine the results with the best fit Semi Analytical Model (SAM)
codes, as defined in M10, to compare the resulting prediction for the luminosity function 
of MW satellite in order to infer a lower limit on the WDM candidate mass.

\section{Simulations and SAMs} 
\label{sec:sims}

In this paper we combine merger trees extracted from very high
resolution N-body simulations,  describing the hierarchical
assembly of a MW-like halo, with SAM techniques, to predict the
relationship between the dark matter (sub)haloes and observable
galaxy properties, allowing us to make a direct and detailed
comparison with observational data from SDSS.
We perform LCDM simulations and analyze them (including the definition
of DM haloes and the reconstruction of their detailed merger tree)
using the same tools described in full detail in M10 and we refer the
reader to that paper for a more detailed discussion. In particular,
for this work we select one specific halo (namely the G1 halo in M10)
and we resimulate it for a suite of WDM models with particle masses
$m_w=10, 5, 2, 1$ keV. The dark matter particle mass is
$m_{d} = 4.16 \times 10^5 \hMsun$, and gravitational softening of 
355 $h^{-1}$ pc, with $\approx 3 \times 10^6$ particles within the virial radius.

To generate initial conditions for WDM, we define a rescaled version
of the CDM power spectrum using a fitting function that approximates
the transfer function associated to the free streaming effect of WDM
particles (Viel \etal 2005). We do not include the effect of a non
zero primordial velocity dispersion for WDM particles in our
simulations, this because even in our more extreme WDM model (1 keV)
the rms of the random velocity component at the starting redshift of
the simulations ($z\approx 40$) is of the order of 2 km sec$^{-1}$.
This is much smaller than the typical velocities induced by the
gravitational potential itself.  Hence we do not expect that
neglecting this random velocity component could alter our results.
In the following, we will not show the results for the $m_w=10$ keV 
model since it turned out to be indistinguishable from LCDM.

In order to predict the expected luminosities of satellite galaxies,
we combine the results of the $N$-body simulations with two
state-of-the-art SAMs, namely the Kang \etal (2005, 2006, see also
Kang 2008) model and \morgana (Monaco \etal 2007, updated in Lo Faro
\etal 2009).
In M10 we studied in full detail the effect of the various physical
processes in shaping the luminosity function of MW satellites. We
determined that suppression of gas infall by a photo-ionizing
background, supernova feedback and tidal destruction are the most
relevant processes responsible for the agreement between theoretical
predictions and data. We characterize our best fit models as
follows. (i) We regulate star formation efficiency in low mass haloes
by shutting off gas cooling in structures with virial temperature
below $10^{4}K$ (due to the inefficiency of H$_{2}$ cooling). (ii) We
suppress hot gas accretion in low mass haloes according to
photo-ionization background, reionization and filtering mass arguments
(Kravtsov \etal 2004). (iii) Stellar feedback is modeled as a function
of halo circular velocity. For the purposes of this work we keep the
parameters fixed at the same values in the best fit models of M10
\footnote{A more detailed and critic discussion of how the variation 
in SAM parameters may affect the final luminosity function can be 
found in M10.}
The two SAMs provide consistent prediction with respect to the dependence
of the LF on the WDM particle mass, and lead to similar
conclusions. For a sake of simplicity, in the following of the paper
we will focus and show only results referring to the Kang model.

\section{Results}
\label{sec:res}

\begin{figure}
\centering{
\psfig{figure=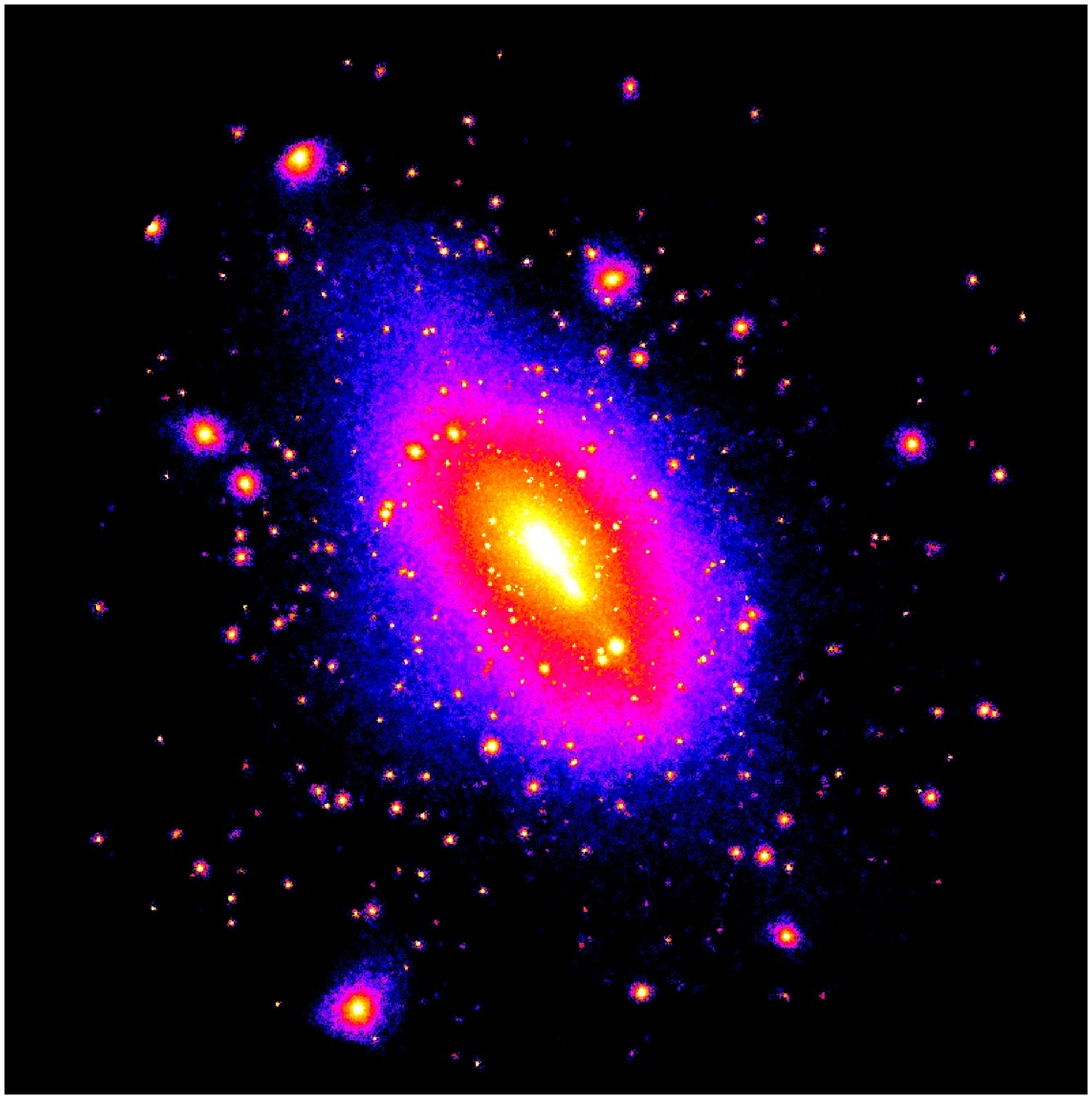,width=0.4\textwidth}
\psfig{figure=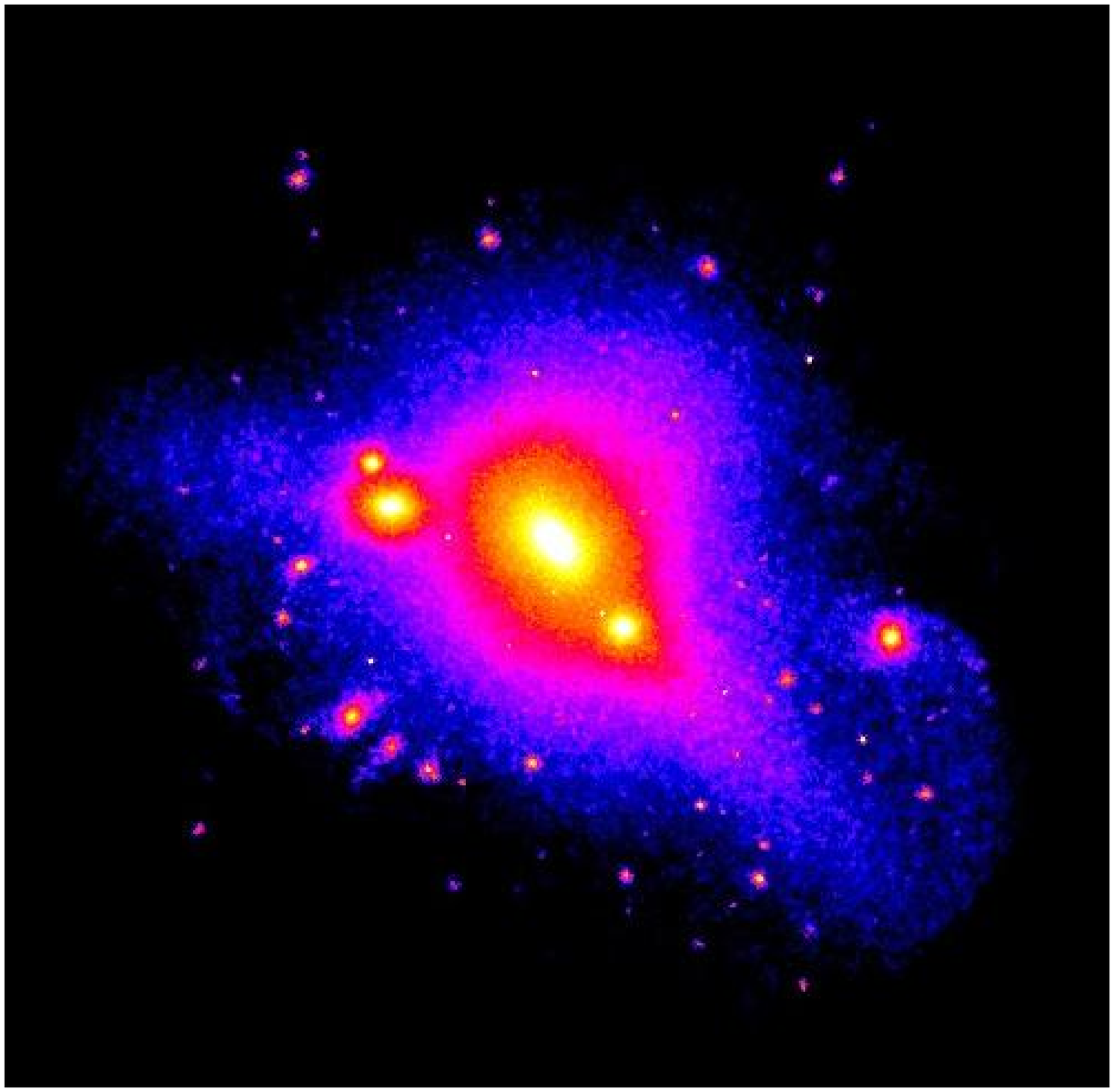,width=0.4\textwidth}
}
\caption{\scriptsize Dark matter density map within 
a sphere of radius $R=360$ kpc. LCDM and LWDM results ($m_w=1$ keV) 
are shown in the upper and lower panel respectively.}
 \label{fig:gal}
\end{figure}

\begin{figure}
\psfig{figure=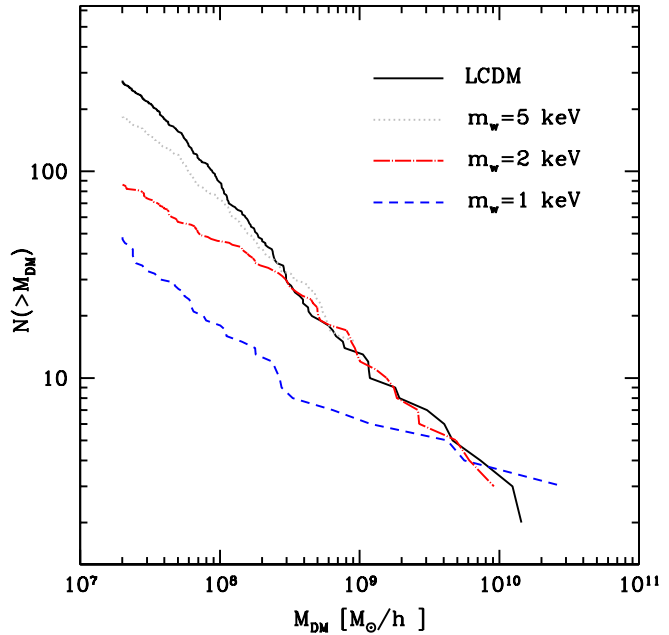,width=0.5\textwidth}
\caption{\scriptsize Subhaloes mass function in LCDM and LWDM
  models. Solid, (gray) dotted, (red) dot-dashed and (blue) dashed lines 
  refer to LCDM and LWDM ($m_{w}$=5, 2 and 1 keV)  respectively.}
\label{fig:MFwdm}
\end{figure}

The effects of WDM are clearly visible in figure \ref{fig:gal}, where
we show the dark matter density map within 360 kpc for the 2 extreme
models: LCDM (upper panel) and LWDM with $m_w=1$ keV (lower panel).
The difference in the number of substructures can be quantified by
looking at the cumulative satellite mass function which is shown in
figure \ref{fig:MFwdm}. In order to identify bound subhaloes we use the 
{\sc ahf}\footnote{The Amiga Halo finder ({\sc ahf}) can be freely downloaded from 
http://www.popia.ft.uam.es/AMIGA} halo finder (Knollmann \& Knebe 2009). 

\begin{figure}
\psfig{figure=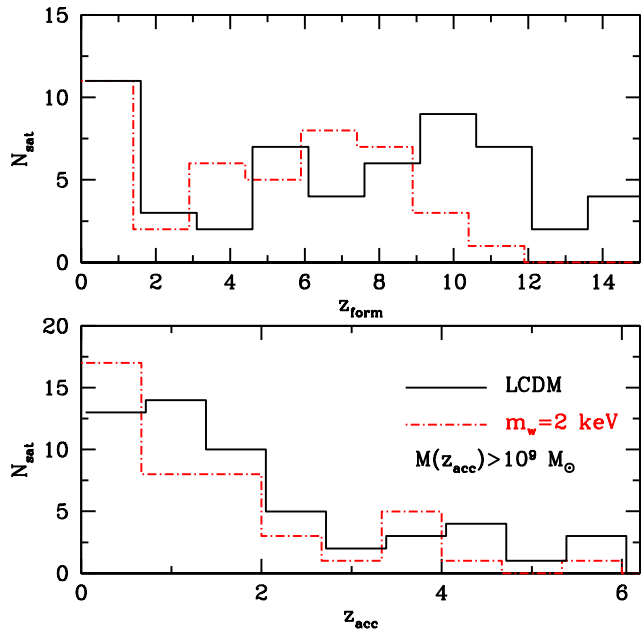,width=0.5\textwidth}
\caption{\scriptsize Lower panel: Distribution of accretion redshifts $z_{acc}$ for subhaloes
with $M(z_{acc})\ge 10^9 \Msun$. Solid (black) line shows results for the LCDM model, while the 
dot-dashed (red) line is for the LWDM with $m_w=2$ keV. Upper panel same as lower panel
but for the formation redshift $z_{form}$, defined as the redshift at which the 
halo virial temperature exceeds $10^4$ K.}
\label{fig:Zac}
\end{figure}

In the WDM scenario not only the total number but also the formation
history of dark matter structures is expected to be different with
respect to LCDM.  Due to the lack of power on small scales (sub)haloes
will both form and be accreted onto the main halo at different
times. This effect could leave a footprint on the luminosity of
  the satellites hosted by the subhaloes. Subhaloes will have
  a different mass at the time of reionization and this will affect
  the extent of gas suppression. Moreover, their accretion redshift
  onto the main halo will be different, with relevant implications for
  the star formation history of the satellites (in the SAMs we
  consider there is no new gas accretion onto subhaloes). Figure
  \ref{fig:Zac} shows the distribution of accretion ($z_{acc}$) and
  formation ($z_{form}$) redshifts, where this latter quantity is
  defined at the time when the halo virial temperature exceeds $10^4$
  K (allowing gas cooling), for the LCDM and LWDM ($m_w=2$ keV)
  models. As expected, gas cooling and star formation start later in
  the WDM scenario, with almost no haloes with $z_{form}=11$ (upper panel).
  At the same time, they seem
  to have (on average) a later accretion time in WDM models with
  respect to LCDM (lower panel). Those two effects act on opposite
  ways in shaping the stellar content of the galaxy hosted by the dark
  matter (sub)halo, and, as we will see next, the net effect on the
  satellite luminosity function is almost negligible.

\begin{figure}
\psfig{figure=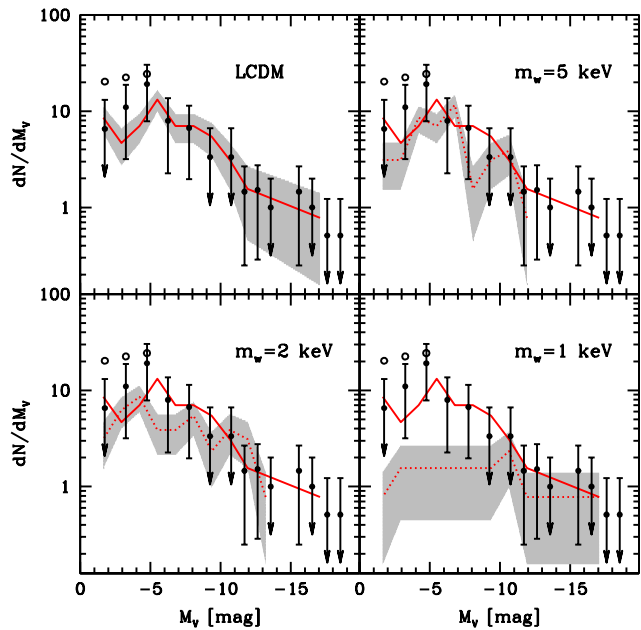,width=0.5\textwidth}
\caption{\scriptsize Milky Way satellites luminosity function in LCDM
and LWDM models. In each panel solid line refers to the prediction
for LCDM cosmology, while dotted lines refer to the predictions for
LWDM realizations, as indicated in the panels.
The shaded area represents the $1\sigma$ 
Poisson scatter around the mean. Observational data are taken from 
Koposov \etal (2008) under the assumption  of two different radial distributions of satellites, NFW-like (solid
circles with error bars) and isothermal (open circles). The arrows
on error bars indicate that there is only one galaxy in that
particular bin, and so the Poisson error is formally 100\%.}
\label{fig:MVwdm}
\end{figure}

Our main results on the MW satellite luminosity function are presented
in figure \ref{fig:MVwdm}. In each panel, we compare the original
results for the G1 halo in the LCDM cosmology (shown as a solid line)
with the predictions corresponding to the different LWDM cosmologies
(dotted lines). We decide to show only G1 in this work since, among
the four haloes presented in M10, it was the one giving the best
agreement with the observed satellite luminosity function.  This makes
the comparison between the observed and simulated LF in WDM models
more straightforward.  It is evident from the figure that the number
density of satellites of different luminosity is reproduced for a WDM
particle $m_w>2$ keV as well as for the standard cosmology. Only when
considering $m_w=1$ keV we detect a significant discrepancy with the
observational data, with only nine predicted satellites fainter than
$M_V=-9$. This result can be used to set a lower limit on on the warm
dark matter particle mass ($m_w>1$ keV), in broad agreement with the
results coming from the analysis of the galaxy power spectrum.

It is worth noting the peculiar behavior of the bright end of the
luminosity function. There is no satellite brighter than $M_V = -12$
in the $m_w= 5$ keV and $m_w = 2$ keV realizations, while the
predictions for the $m_w =1$ keV model are in broad agreement with the
observational data. We checked that this peculiar behavior is not
related to the modeling of the dynamical friction or tidal stripping
in the SAMs, but it is related to the different properties of G1
merger tree in the various LWDM cosmologies. As an effect of the
reduced small-scale power, the statistical properties of DM haloes
change: haloes are on average less concentrated (Eke \etal 2001) and
both the accretion time and the mass of a dark matter halo at that
time are modified with respect to LCDM. Given the small number
statistics associated to the bright end of the satellite LF for a
galaxy-sized halo (less than 4 objects in total), this result is not
totally unexpected. For this reason we stress that the stronger
constraints come from the comparison of the faint end, where the
number statistics are high enough to reveal any significant
modification of the LF with respect to the LCDM result.

Mateo \etal (1993) and Strigari \etal (2008, see also Walker \etal 2009) 
have noted that all satellites with luminosity between $10^3$ and
$10^7 L_{\odot}$ have a common mass of $\sim 10^7 M_{\odot}$ within a
radius of 300pc ($M_{0.3}$). Moreover, within this inner region, all
objects result to be dark matter dominated. Macci\`o, Kang \& Moore
(2009, see also Li \etal 2009b, Okamoto \& Frenk 2009) showed that this
relation is not totally unexpected in a LCDM Universe. Given the
expected modification in the properties of DM haloes as a function of
redshift in a WDM model, it is interesting to check if our simulated
satellites are still able to reproduce the observed normalization and
small scatter of the $M_{0.3}-L$ relation.

\begin{figure}
\psfig{figure=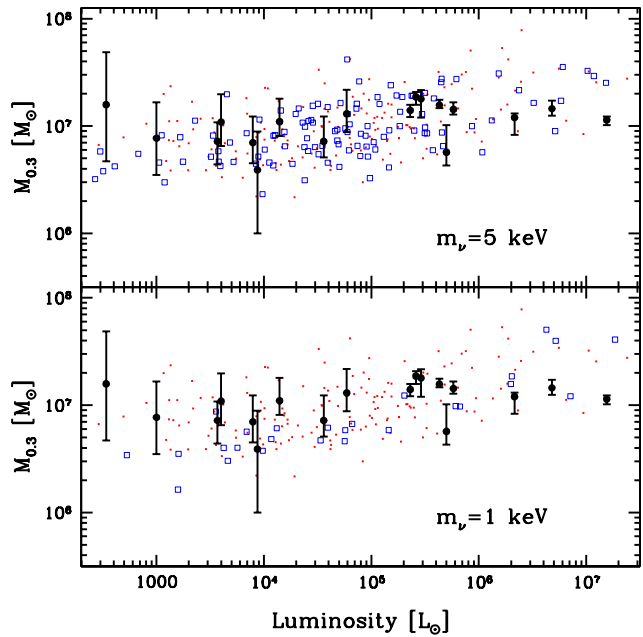,width=0.5\textwidth}
\caption{\scriptsize Mass within 300 pc versus luminosity.  Red dots
  show results from LCDM numerical simulations. The (blue) squares
  show results for WDM satellites for a $m_{\nu}$ mass of 5 keV and 1
  keV in the upper and lower panel respectively.  Black points with
  error bars are the observational results from Strigari \etal (2008).}
\label{fig:M03wdm}
\end{figure}

Satellite luminosities are direct outputs of the SAM, while for
computing $M_{0.3}$ we used the same approach detailed in Macci\`o,
Kang \& Moore (2009). At time of accretion of each satellite we
compute the parameters ($\rs$ and $\delta_c$) that describe its
density profile assumed to be NFW (Navarro \etal 1997, see 
Macci\`o \etal 2008 for more details).
The present value of $M_{0.3}$ is then computed by integrating the 
density profile, under the assumption that 
$\rs$ and $\delta_c$ do not evolve with time.
Figure \ref{fig:M03wdm} shows the distribution of the the mass within 300 pc
as a function of the satellite luminosity. The upper panel presents
results for a WDM with $m_w=5$ keV (blue open squares), which turn out
to be indistinguishable from the LCDM ones (red points) and hence in
good agreement with the observational findings of Strigari \etal 2008.
The lower panel shows results for our most extreme WDM model ($m_w=1$
keV): in this case simulated satellites suggest a stronger correlation
between their inner mass and luminosity which is in (slight)
disagreement with the flat distribution of the observational
data. Nevertheless, this discrepancy is not as strong as that seen for
the luminosity function and it does not allow us to reject the $m_w=1$
keV model. We should also point out that our determination of
$M_{0.3}$ can be partially affected by neglecting
primordial thermal velocities that would create a core in the density
profile (Strigari \etal 2006). The expected size of the core is
100 and 8 pc for the $m_w=1$, $m_w=5$ keV models respectively 
(Strigari \etal 2006), thus for $m_w=5$ keV our approach for computing 
$M_{0.3}$could be not fully self-consisted.

\section{Discussion and conclusions}
\label{sec:concl}

In this paper we compare the most recent observational data on the
luminosity function of MW satellites with a suite of $N$-body
simulations combined with SAM techniques, with the aim of constraining
a possible ``warm'' nature for DM particle. In particular, we want 
to make use of the new tight observational constraints on the faint-end, 
which have already been proved to be of fundamental importance to test SAMs 
in the light of the problem of MW missing satellites (see M10)

We perform a series of $N$-body simulations with different choices for
the mass of the warm dark matter particle ($m_w$), and we combine them
with Semi-Analytical Models (SAMs) for galaxy formation, using the
best fit solution found in M10 for the parameters describing the SAMs.
We then compare the resulting statistics of MW satellites with both
the predictions for a LCDM Universe and observational data.

Our results show that we are indeed able to put a lower limit on $m_w$
(1 keV), which is in agreement with previous determinations based on
the galaxy power spectrum and Lyman-$\alpha$ forest
observations (Viel \etal 2008, Boyarsky \etal 2009a).
It is worth noting that our limits
are less tight with respect to the previous determinations, as a
number of caveats have to be taken into account. First, only the
$m_w=1$ keV realization really fails on reproducing the MW satellites
luminosity function (with our ``standard'' choice of SAM
parameters). Using a $m_w$ value as small as 2 keV, we still predict a
luminosity function which is almost indistinguishable from the
concordance LCDM cosmology. This is due to the fact that the for
$m_w>1$ keV, the small-scale power modifications do not affect
considerably the substructures responsible for the formation of the
bulk of visible MW satellites, showing an intrinsic limit of this
approach with the current observational data. Moreover, the faint-end
of the LF, which provides us the strongest statistical constraints, is
sensitive to the details of the description of baryonic physics
included in the SAMs, and in particular to the strength of stellar
feedback in low-mass halo (see M10). Therefore only models that
drastically failed in reproducing the data (as $m_w=1$ keV) can be
realistically ruled out.
In the present work we only considered a very simple WDM model;
it is worth notice that there are more complex and more physically 
motivated models discussed in the literature (e.g. warm+cold dark matter,
Boyarsky \etal 2009b or composite dark matter Khlopov 2006, Khlopov \& Kouvaris 2008).
These scenarios deserve special studies and can provide new, interesting
hints on the nature of dark matter. 

This paper is mainly a ground test on the feasibility of our approach. 
New data coming from the most recent multiwavelength surveys like the
Panoramic Survey Telescope and Rapid Response System
(Pan-STARRS)\footnote{http://www.ps1sc.org/index.htm} will provide a
better determination of the MW satellite properties and then an optimal
data-set to increase our knowledge about the nature of dark matter.

\section*{Acknowledgments}

The authors are extremely grateful to X. Kang for sharing the update
version of his SAM. We also thank A. Boyarsky, H. de Vega, M. Khlopov 
and O. Ruchayskiy
for helpful discussions on this manuscript and the referee, Mario Mateo, for 
constructive comments that improved the presentation of the paper.
Numerical simulations were performed on the PIA
and on PanStarrs2 clusters of the Max-Planck-Institut f\"ur Astronomie
at the Rechenzentrum in Garching.


\end{document}